\documentclass[10pt,conference]{llncs2e/llncs}
\usepackage{listings}
\usepackage{amsfonts,amscd}
\usepackage{amssymb}
\usepackage{mathtools}
\usepackage{comment}
\usepackage{enumerate}
\usepackage{booktabs}
\usepackage{float}
\usepackage{verbatim}
\usepackage{lineno,hyperref}
\modulolinenumbers[5]
\usepackage{float}
\usepackage[]{graphicx}
\usepackage{stackengine}
\usepackage{soul}
\usepackage{xcolor}
\usepackage{quantikz}
\usepackage{multirow}
\usepackage{braket}
\usepackage{physics}
\usepackage{arydshln}
\usepackage{graphicx}
\usepackage{subcaption}
\RequirePackage{algorithm}
\RequirePackage[noend]{algpseudocode}
\usepackage{todonotes}
\usepackage{booktabs}
\usepackage{dsfont}
\def\unit{\mathds{1}}
\usepackage{comment}
\usepackage{tabularx}

\let\subsectionautorefname\sectionautorefname

\spnewtheorem{ex}{Example}{\bf}{\itshape}

\makeatletter
\renewcommand*\env@matrix[1][c]{\hskip -\arraycolsep
  \let\@ifnextchar\new@ifnextchar
  \array{*\c@MaxMatrixCols #1}}
\makeatother

\graphicspath{{fig/}}

\begin{document}

\newcommand\concept[1]{\textit{#1}}
\renewcommand\sectionautorefname{Section}
\renewcommand{\subsectionautorefname}{Section}

\title{
Fast equivalence checking of quantum circuits of Clifford gates
}

\pagestyle{plain}

\author{Dimitrios Thanos \and 
Tim Coopmans \and 
Alfons Laarman\
}
\institute{Leiden Institute of Advanced Computer Science (LIACS), Leiden University,  Leiden  2333 CA, The Netherlands}

\maketitle

\begin{abstract}
Checking whether two quantum circuits are equivalent is important for the design and optimization of quantum-computer applications with real-world devices.
We consider quantum circuits consisting of Clifford gates, a practically-relevant subset of all quantum operations which is large enough to exhibit quantum features such as entanglement and forms the basis of, for example, quantum-error correction and many quantum-network applications.
We present a deterministic algorithm that is based on a folklore mathematical result and demonstrate that it is capable of outperforming previously considered state-of-the-art method.
In particular, given two Clifford circuits as sequences of single- and two-qubit Clifford gates, the algorithm checks their equivalence in $O(n \cdot m)$ time in the number of qubits $n$ and number of elementary Clifford gates $m$.
Using the performant Stim simulator as backend, our implementation checks equivalence of quantum circuits with 1000 qubits (and a circuit depth of 10.000 gates) in $\sim$22 seconds and circuits with 100.000 qubits (depth 10) in $\sim$15 minutes, outperforming the existing SAT-based and path-integral based approaches by orders of magnitude.
This approach shows that the correctness of application-relevant subsets of quantum operations can be verified up to large circuits in practice.

\end{abstract}

\section{Introduction}

Quantum computing promises to perform classically intractable tasks for a wide range of applications~\cite{nielsen2000quantum,montanaro2016quantumalgorithms}.
While we are entering the era of Noisy Intermediate-Scale Quantum computing~\cite{preskill2018quantum}, the high noise levels necessitate precise compilation of textbook quantum circuits onto real-world devices, which can only handle shallow-depth circuits and have various constraints (connectivity, topology, native gate sets, etc.)~\cite{finigan2018qubit,corcoles2019challenges}.
A crucial part of the design and optimization over quantum circuits is \emph{verifying} whether two quantum circuits, each presented by a classical description, implement the same quantum operation, i.e. checking equivalence of quantum circuits.

Correctness verification is a well-studied field in the classical domain~\cite{Novikov03,Ganai02,Kuehlmann04} but unfortunately not all methods directly carry over to quantum computing because the state of $n$ quantum bits is generally represented as $2^n$ complex values~\cite{nielsen2000quantum}.
Due to the reversibility of quantum circuits,
verifying equivalence of circuits $C_1, C_2$ is reducible to checking if the circuit $C_1\cdot C_2^{-1}$, i.e., $C_1$ followed by the inverse of $C_2$, is equivalent to the identity circuit, i.e., a circuit that implements an operator that does not modify the inputs.
Exact and approximate identity checking, i.e., determining whether the circuit is close to the identity circuit, fall in quantum complexity classes which are analogs of NP~\cite{bookatz2012qma,adleman1997quantum} (NQP~\cite{tanaka2010exact} and QMA~\cite{janzing2005non,ji2009non}, respectively).
Thus we should not hope for efficient algorithms in general.

Existing deterministic methods analyzing circuits consisting of only quantum gates as quantum operations (no quantum measurements) are based on encoding as Boolean satisfiability instances~\cite{berent2022towards} (also \cite{wille2013compact,yamashita2010fast} for restricted circuits), satisfiability modulo theories~\cite{bauermarquart2023symqv}, path-sums~\cite{amy2018towards,amy2019thesis}, rewrite rules~\cite{peham2022equivalencecheckingquantum,Duncan20,Wetering20}, and on various flavors of decision diagrams, including QMDD~\cite{burgholzer2020improved,viamontes2007equivalence,niemann2014equivalence,burgholzer2020advanced}, LIMDD~\cite{limdd,vinkhuijzen2023efficient}, Tensor-DD~\cite{hong2021approximate}, BDD~\cite{wei2022accurate,chen2022partial} and others~\cite{wang2008xqdd,yamashita2008ddmf}.
In addition, some probabilistic methods are known~\cite{burgholzer2021random,linden2021lightweight}.

In this paper, we focus on exact equivalence checking of two (classical descriptions of) circuits with Clifford gates only, a subset of all quantum gates which is ubiquitous to quantum computing and is highly relevant for quantum error correction~\cite{gottesman1997stabilizer,terhal2015quantum} and quantum networking applications~\cite{hein2006entanglement}.
For exact identity checking of Clifford circuits, a reduction to satisfiability was presented by~\cite{berent2022towards}, in a tool called QuSAT, and an approach based on path-sums in the Feynman tool~\cite{amy2018towards}.
For approximate identity check, a polynomial-time algorithm exists~\cite{arunachalam2022parameterized} whose runtime scales with the accuracy of the approximation (a polynomial in the number of qubits).

We demonstrate that a folklore characterization of equivalence of general circuits translates into an $O(m \cdot n)$-time deterministic algorithm for exact equivalence checking of Clifford circuits, with $n$ the number of qubits and $m$ the total number of elementary Clifford gates in the two circuits.
The algorithm (many-one) reduces the equivalence check to circuit simulation which can be done efficiently~\cite{gottesman1998heisenberg,aaronson2008improved}, 
in the particular case of equivalence checking of Clifford circuits.

We empirically evaluate the algorithm by using the performant Clifford-circuit simulator Stim~\cite{Gidney2021stimfaststabilizer}, reaching circuit depths of 1000 qubits and 10.000 elementary Clifford gates in less than a minute, and 100.000 qubits for depth-10 circuits in approximately 15 minutes, outperforming the state-of-the-art SAT-based and path-sum approaches by orders of magnitude.
Our open-source implementation can be found on \cite{githubrepo}.

We emphasize that the task in this work is equivalence checking given a white-box \emph{classical} descriptions of the quantum circuit, as opposed to the different task where one is given access to a \emph{quantum} computer which performs the quantum circuit as black box~\cite{montanaro2013survey}.
For Clifford circuits, specifically see~\cite{brakerski2021unitary} and \cite{linden2021lightweight}.

In \autoref{sec:preliminaries}, we provide the necessary background to quantum computing and a simple example of applying the algorithm for comparing two equivalent circuits.
We state the theorem explicitly and give the resulting algorithm in \autoref{sec:theor_alg}.
In \autoref{sec:experiments}, we empirically evaluate our implementation, using the Stim simulator as a backend.
We conclude in \autoref{sec:conclusions}.

\section{Preliminaries \label{sec:preliminaries}}

We briefly introduce relevant quantum computing concepts and refer to \cite{nielsen2000quantum} for a more elaborate introduction.

\subsection{Quantum circuits and fundamental concepts}\label{quantum circuits}
Classical circuits are limited to bits, which take values 0 and 1.
In contrast, the state of a quantum bit or qubit can be expressed as a complex-valued 2-vector of unit norm.
Examples of single-qubit states are $\begin{bmatrix} \frac{1}{\sqrt{2}} \\ \frac{1}{\sqrt{2}} \end{bmatrix}$ and
$\begin{bmatrix} \frac{1}{\sqrt{5}} \\ \frac{2i}{\sqrt{5}} \end{bmatrix}$ where $i$ is the imaginary unit ($i^2 = -1$).
Two possible quantum states are the computational-basis states $\begin{bmatrix} 1 \\ 0\end{bmatrix}$ and $\begin{bmatrix} 0 \\ 1 \end{bmatrix}$, usually denoted in Dirac notation as $\ket{0}$ and $\ket{1}$.
Thus, we can rewrite the two examples from before as
$\frac{1}{\sqrt{2}} (\ket{0} + \ket{1}) = \begin{bmatrix} \frac{1}{\sqrt{2}} \\ \frac{1}{\sqrt{2}} \end{bmatrix}$
and $\frac{1}{\sqrt{5}} (\ket{0} + 2i \ket{1}) = \begin{bmatrix} \frac{1}{\sqrt{5}} \\ \frac{2i}{\sqrt{5}} \end{bmatrix}$.
More generally, we can write an arbitrary single-qubit state $\ket{\phi}=\begin{bmatrix} \alpha_0 \\ \alpha_1 \end{bmatrix} = \alpha_0\ket{0}+\alpha_1\ket{1}$ where the complex numbers $\alpha_i$ satisfy $|\alpha_0|^2 + |\alpha_1|^2 = 1$.
Here, $|z|$ denotes the modulus of the complex number $z$: when writing $z = a + b\cdot i$  for real numbers $a, b$, the modulus equals $|z| = \sqrt{a^2 + b^2}$ and can also be defined through the complex conjugate $z^* = a - b\cdot i$ as $|z| = \sqrt{z \cdot z^*}$.

Two single-qubit quantum states $\ket{\phi}, \ket{\psi}$ are combined into a two-qubit state $\ket{\phi} \otimes \ket{\psi}$, where $\otimes$ denotes the tensor product (Kronecker product) from linear algebra.
In general, an $n$-qubit state is a complex vector of $2^n$ entries and can be written in Dirac notation as $\sum_{x\in \{0, 1\}^n} \alpha_x \ket{x}$, where $\ket{x}$ are defined as e.g. $\ket{0010} = \ket{0} \otimes \ket{0} \otimes \ket{1} \otimes \ket{0}$.
Here, the complex values $\alpha_x$ should satisfy $\sum_{x\in \{0, 1\}^n} |\alpha_x|^2 = 1$, i.e. the norm of the vector representing the quantum state equals 1.
Examples of two-qubit states are $\ket{00}$ and $\frac{1}{\sqrt{6}}(\ket{00} + i\ket{01} - 2\ket{11})$.
Any $(n_A+n_B)$-qubit quantum state $\ket{\phi}$ that cannot be written as a product state $\ket{\phi_A}\otimes \ket{\phi_B}$, with $\ket{\phi_A}$ ($\ket{\phi_B}$) a state of $n_A$ ($n_B$) qubits, is called entangled, e.g. $\frac{1}{\sqrt{2}}(\ket{00} + \ket{11})$.

There are two main operations on quantum states in the usual circuit model: quantum gates and quantum measurements.
We will only use gates here. 
A quantum gate on $n$ qubits is a unitary operator that is represented by a $2^n$ by $2^n$ unitary matrix $U$.
(Unitarity, defined below, ensures that the operator is reversible and norm-preserving.)
A quantum state $\ket{\phi}$ is updated by a unitary matrix as $U\cdot \ket{\phi}$ where $\cdot$ denotes matrix-vector multiplication.
As example, consider the following single-qubit gates:
\[
\text{Hadamard: } H = 
\frac{1}{\sqrt{2}} \begin{bmatrix}[r]
    1 & 1 \\
    1 & -1
    \end{bmatrix}
\qquad
\text{Phase gate: } S = 
\begin{bmatrix}
    1 & 0 \\
    0 & i
    \end{bmatrix}
.
\]
Applying the Hadamard gate to the state $\ket{0}$ for example, we obtain
\[
   H\cdot\ket{0} = \frac{1}{\sqrt{2}} \begin{bmatrix}[r]
    1 & 1 \\
    1 & -1
    \end{bmatrix} \cdot \begin{bmatrix}
    1 \\
    0
    \end{bmatrix} = \frac{1}{\sqrt{2}} \begin{bmatrix}
    1 \\
    1
    \end{bmatrix} = \frac{\ket{0} + \ket{1}}{\sqrt{2}}
\]
and similarly one can compute $ S \ket{1} = i \ket{1}$.

A quantum gate $U$ is a unitary matrix, which means $U\cdot U^{\dagger} = U^{\dagger} \cdot U = \unit_{2^n}$, where $\unit_{2^n}$ is the identity matrix on vectors of $2^n$ entries (i.e. the matrix that has the property $\unit_{2^n}\cdot \vec{v} = \vec{v}$ for each vector $\vec{v}$ of $2^n$ complex numbers) and the adjoint operator $(.)^{\dagger}$ means transposing the matrix and replacing each matrix entry by its complex conjugate.  For example, the Hadamard gate and phase gate have adjoint operators
\[
H^{\dagger} = 
\frac{1}{\sqrt{2}} \begin{bmatrix}[r]
    1 & 1 \\
    1 & -1
    \end{bmatrix}
=H
\qquad
S^{\dagger} = 
\begin{bmatrix}[r]
    1 & 0 \\
    0 & -i
    \end{bmatrix}
.
\]
It is not hard to check that indeed $H^{\dagger} \cdot H = S^{\dagger} \cdot S = \unit_2 = \begin{bmatrix} 1 & 0 \\ 0 & 1 \end{bmatrix}$.
Applying an $n$-qubit gate $A$ to the first part of an $(n+m)$-qubit quantum state $\ket{\phi}$ is done by tensoring with the identity, i.e. $A\otimes \unit_{2^m}$ is applied to the entire state $\ket{\phi}$.

Another notion which we will use later is the bra $\bra{\phi} = (\ket{\phi})^{\dagger}$ and the inner product $\langle \phi | \cdot | \psi\rangle =\langle \phi | \psi\rangle = \sum_{x\in\{0, 1\}^n} a_x^* \cdot b_x$ for $\ket{\phi} = \sum_{x\in\{0, 1\}^n} a_x \ket{x}$ and $\ket{\psi} = \sum_{x\in\{0, 1\}^n} b_x \ket{x}$.
Observe that state normalization implies that $\braket{\phi} = 1$.

A quantum circuit is composed of qubits represented by horizontal lines (wires) and quantum gates represented by boxes, with each gate acting on one or more qubits. The input state is represented to the left of the wires of the circuit (typically $\ket{0}^{\otimes n}$), and the output state is obtained after the sequential application of the circuit's gates.
 These gates, which are typically unitary operators of small size, can be combined with matrix multiplication into a single unitary operator describing the entire circuit as a single operator. See \autoref{ex1} for an explicit calculation of a quantum circuit's output state. A quantum algorithm is a uniform family of quantum circuits (like in circuit complexity~\cite{arora2009computational}).

A special but important class of quantum circuits are the Clifford circuits.
Any Clifford gate can be written as a Clifford circuit\footnote{When we say `circuit', we mean the sequence of quantum gates, i.e. without the input state, e.g., $\ket{0}^{\otimes n}$.} consisting only of three elementary Clifford gates: $H, S$ and $CNOT$. 
The controlled not ($CNOT$) gate acts on two qubits: The first is called the ``control'' and the second one the ``target'' qubit. It is symbolized by a vertical line connecting the two qubits, with a dot representing the control qubit and $\oplus$ representing the target (see \autoref{ex1}).
The $CNOT$ gate and its inverted counterpart $NOTC$ (see circuit~B in \autoref{ex1}) are defined as the following two-qubit operators: $$
CNOT=\begin{bmatrix}
1 & 0 & 0 & 0 \\
0 & 1 & 0 & 0 \\
0 & 0 & 0 & 1 \\
0 & 0 & 1 & 0 \\
\end{bmatrix},
~~~~~~
NOTC = 
\begin{bmatrix}
1 & 0 & 0 & 0 \\
0 & 0 & 0 & 1 \\
0 & 0 & 1 & 0 \\
0 & 1 & 0 & 0 \\
\end{bmatrix}. $$
One of the significant features of Clifford circuits is that they can be simulated in polynomial time in the number of qubits and number of elementary Clifford gates by classical computers, as shown by the Gottesman-Knill theorem \cite{gottesman1998heisenberg} (see also \autoref{sec:pauli}). In addition, with the usual input $\ket{0}^{\otimes n}$, Clifford circuits can generate various entangled states, and become universal --- meaning they can approximate any quantum circuit --- if non-Clifford gates are added to the gate set \cite{Universal}.\\

\begin{ex}
	\label{ex1}
\noindent We provide an example for calculating the output states of the two circuits $A$ and $B$ below. This example uses Hadamard gates and CNOT gates.\\
\begin{center}
\begin{minipage}[t]{0.45\textwidth}
\centering
{A)}
\begin{quantikz}
    \lstick{} & \gate{H} & \ctrl{1} & \gate{H} & \qw \\
    \lstick{} & \gate{H} & \targ{} & \gate{H} & \qw \\
\end{quantikz}
\end{minipage}
\hfill
\begin{minipage}[t]{0.45\textwidth}
\centering
{B)}
\begin{quantikz}
    \lstick{} & \lstick{} & \targ{1} & \qw &\rstick{} \\
    \lstick{} & \lstick{} & \ctrl{-1} & \qw & \rstick{} \\
\end{quantikz}
\end{minipage}
\end{center}
In this example, we assume here that the initial state on each qubit is $\ket{0}$.

\begin{enumerate}[A)]
\item We start by making the calculations for circuit A. For a Hadamard gate applied on the first qubit we get: 
$$
H\ket{0} = \frac{1}{\sqrt{2}}\begin{bmatrix}[r]
1 & 1 \\
1 & -1 \\
\end{bmatrix}
\begin{bmatrix}
1 \\
0 \\
\end{bmatrix}
= \frac{1}{\sqrt{2}}\begin{bmatrix}[r]
1 \cdot 1 + 1 \cdot 0 \\
1 \cdot 1 + (-1) \cdot 0 \\
\end{bmatrix}
= \frac{1}{\sqrt{2}}\begin{bmatrix}
1 \\
1 \\
\end{bmatrix}
=
		\frac{1}{\sqrt{2}}\left( \ket{0} + \ket{1} \right)
$$

The result of applying a Hadamard gate on the second qubit will be the same. Since the two qubits are independent we can calculate the state of this system by tensoring the two states:

$$
\begin{aligned}
&\left(\frac{1}{\sqrt{2}} (|0\rangle + |1\rangle)\right) \otimes \left(\frac{1}{\sqrt{2}} (|0\rangle + |1\rangle)\right) \\
&= \frac{1}{\sqrt{2}} \frac{1}{\sqrt{2}} (|0\rangle + |1\rangle) \otimes (|0\rangle + |1\rangle) \\
&= \frac{1}{2} (|0\rangle \otimes |0\rangle + |0\rangle \otimes |1\rangle + |1\rangle \otimes |0\rangle + |1\rangle \otimes |1\rangle).
\\
	&=
\frac{1}{2} (|00\rangle + |01\rangle + |10\rangle + |11\rangle)
.
\end{aligned}
$$

In matrix notation this would be:

$$
\frac{1}{2} \begin{bmatrix}
1 \\ 0 \\ 0 \\ 0
\end{bmatrix} +
\frac{1}{2} \begin{bmatrix}
0 \\ 1 \\ 0 \\ 0
\end{bmatrix} +
\frac{1}{2} \begin{bmatrix}
0 \\ 0 \\ 1 \\ 0
\end{bmatrix} +
\frac{1}{2} \begin{bmatrix}
0 \\ 0 \\ 0 \\ 1
\end{bmatrix} =
\frac{1}{2} \begin{bmatrix}
1 \\ 1 \\ 1 \\ 1
\end{bmatrix}.
$$

Now we will calculate how this state is transformed when we apply a CNOT gate where the first qubit is the control qubit and the second one is the target qubit:

$$
CNOT \cdot \frac{1}{2} \begin{bmatrix}
1 \\
1 \\
1 \\
1
\end{bmatrix} = \begin{bmatrix}
1 & 0 & 0 & 0 \\
0 & 1 & 0 & 0 \\
0 & 0 & 0 & 1 \\
0 & 0 & 1 & 0 \\
\end{bmatrix} \cdot \frac{1}{2}  \begin{bmatrix}
1 \\
1 \\
1 \\
1
\end{bmatrix} =  \frac{1}{2}   \begin{bmatrix}
1 \\
1 \\
1 \\
1
\end{bmatrix}
$$

So the state remains the same. The CNOT is interpreted as follows: whenever $b_1=1$ flip the second input of $\ket{b_1 b_2}$. So we can directly calculate $\frac{1}{2}(\ket{00} + \ket{01} + \ket{10} + \ket{11})$ which becomes $\frac{1}{2}(\ket{00} + \ket{01} + \ket{11} + \ket{10})$ without explicitly performing the matrix multiplication. We observe that, as expected, this method returns an identical state. Finally, we apply the two remaining Hadamard gates to the state $\frac{1}{2}(\ket{00} + \ket{01} + \ket{10} + \ket{11})$.

$$
(H \otimes H)\left(\frac{1}{2}\begin{bmatrix}
1 \\
1 \\
1 \\
1 \\
\end{bmatrix}\right) = \frac{1}{4}\begin{bmatrix}
1 & 1 & 1 & 1 \\
1 & -1 & 1 & -1 \\
1 & 1 & -1 & -1 \\
1 & -1 & -1 & 1 \\
\end{bmatrix} \begin{bmatrix}
1 \\
1 \\
1 \\
1 \\
\end{bmatrix} = \begin{bmatrix}
1 \\
0 \\
0 \\
0 \\
\end{bmatrix} = \ket{00}
$$

Therefore, applying Hadamard gates to each qubit of the circuit starting with the initial state $\frac{1}{2}(\ket{00} + \ket{01} + \ket{10} + \ket{11})$ results to the state $\ket{00}$.
\\

\item Now for the much simpler circuit B, we start again by both qubits in the state $\ket{0}$. Following the same rules that we applied for the CNOT of circuit B, the resulting state will also be $\ket{00}$.

\end{enumerate}

Note that in the examples above, we have applied the gates in steps. Generally, this in not necessary, as one can combine the small unitary transformations (the gates) into a single unitary transformation. This, typically large, unitary operator will have the same effect as applying the gates in steps.
For instance, for circuit A, we obtain the following operator $U_A$, which is equal to the $U_B$ unitary ($NOTC$) for circuit $B$:
$$
U_A =
(H\otimes H)\cdot CNOT \cdot  (H\otimes H) =  
 \begin{bmatrix}
1 & 0 & 0 & 0 \\
0 & 0 & 0 & 1 \\
0 & 0 & 1 & 0 \\
0 & 1 & 0 & 0 \\
\end{bmatrix} = NOTC = U_B.
$$

\end{ex}

\subsection{Stabilizer states \label{sec:pauli}}

The Pauli gates are defined as follows:

\begin{equation*}
I = \unit_2 = \begin{bmatrix}1 & 0 \\ 0 & 1\end{bmatrix}, \quad
X = \begin{bmatrix}0 & 1 \\ 1 & 0\end{bmatrix}, \quad
Y = \begin{bmatrix}[r]0 & -i \\ i & 0\end{bmatrix}, \quad
Z = \begin{bmatrix}[r]1 & 0 \\ 0 & -1\end{bmatrix}.
\end{equation*}

\noindent
The $n$-qubit Pauli group $\mathcal{P} _{n}$ is the set $\{\alpha P \mid \alpha \in \{\pm 1, \pm i\}, P \in  \textsc{Pauli}_n\}$ where $\textsc{Pauli}_n$ is the tensor product of $n$ Pauli operators (a ``Pauli string").
For example, we have $X\otimes Z \otimes Y \otimes Y \in \textsc{Pauli}_4$ 
and $-i X\otimes Z \otimes Y \otimes Y \in \mathcal{P}_4$.
The Pauli group forms a group under matrix multiplication.
Any two elements $P_k,P_l\in \mathcal{P}_n$ either commute or anti-commute: either $P_k\cdot P_l = P_l \cdot P_k$ or $P_k\cdot P_l = - P_l \cdot P_k$. Finally, we can give an alternative, equivalent definition of the Clifford group in terms of Pauli matrices: the Clifford group is the set of unitary operators that leave the Pauli group fixed when acting on it by conjugation, i.e. all the $2^n\times 2^n$ unitary matrices $V$ such that $ V P V^{\dagger }\in \mathcal{P} _{n}$ for all $P\in \mathcal{P}_{n}$. 

We will now lay out the stabilizer formalism for efficient classical simulation of Clifford circuits~\cite{gottesman1998heisenberg,aaronson2008improved}.
A unitary operator $U$ \textit{stabilizes} a quantum state if $U\ket{\phi}=\ket{\phi}$.
The so-called stabilizer states form a strict subset of all quantum states which can be described as stabilized by maximal commutative subgroups of the Pauli group using $n$ elements of $\mathcal{P}_n$.
For example, the state $|0\rangle$ is stabilized by the group $\{I,Z\}$ because $I\ket{0} = \ket{0}$ and $Z\ket{0} = \ket{0}$.
Another example is the state $\ket{+} = \frac{1}{\sqrt{2}}(\ket{0} + \ket{1})$, which is stabilized by $\{I,X\}$.
If $\ket{\phi}$ and $\ket{\psi}$ are stabilizer states with stabilizer groups $G, H$, respectively, then $\ket{\phi} \otimes \ket{\psi}$ is also a stabilizer state with stabilizer group $\{g \otimes h \mid g\in G, h\in H\}$.
For example, the state $\ket{11} = \ket{1}\otimes \ket{1}$ is stabilized by the group $\{I\otimes I, -I\otimes Z, -Z\otimes I, Z\otimes Z\}$.
Some stabilizer states are entangled, such as $\frac{1}{\sqrt{2}}\left(\ket{00} + \ket{11}\right)$, which is stabilized by $\{I\otimes I, X\otimes X, -Y\otimes Y, Z\otimes Z\}$.

Maximal commutative subgroups of the Pauli group only have a single quantum state they stabilize~\cite{nielsen2000quantum}; thus, we can \emph{represent} any stabilizer state by its stabilizer group, instead of by providing its description as a vector of $2^n$ complex numbers.
The stabilizer group of an $n$-qubit stabilizer state has $2^n$ elements, so storing all of those would not yield a succinct description of the state.
However, the stabilizer group can be succinctly represented by the generator set of the stabilizer group, which only has $n$ elements $\in\mathcal{P}_n$.
For example, the set $E = \{ -Z\otimes I, -I\otimes Z  \}$ is a set of generators for the stabilizer group $G = \{I\otimes I, -I\otimes Z, -Z \otimes I, Z\otimes Z\}$ of the state $\ket{11}$ because each element of $G$ can be written as a product of elements of $E$. 
Since there are four Pauli gates, we can represent Pauli gate using $\log_2(4) = 2$ bits.
Furthermore, one can show that each element of a stabilizer group is of the form $\pm P_1 \otimes \dots \otimes P_n$ with $P_j$ a Pauli gate; thus, $2n + 1$ bits are needed to represent an element of an $n$-qubit stabilizer group~\cite{Garcia2014,aaronson2008improved} ($2n$ for the Pauli gates in the Pauli string and the $1$ bit for the prefactor $\pm$).
Therefore, by this method, only $n\cdot (2n + 1) = 2n^2+n$ bits are required for the description of a quantum state that can be generated by Clifford circuits while a naive description would require $2^n$ complex numbers. To emphasize the quadratic structure of the generator set, we will write the generators in the so-called \concept{tableau form}, e.g. for the stabilizer generators of $\ket{11}$:
\[
E = \begin{Bmatrix}
- Z\otimes I \\
- I\otimes Z
\end{Bmatrix} \cong
\begin{Bmatrix}
\phantom- Z\otimes Z \\
- I\otimes Z
\end{Bmatrix}
\]

Here, the tableau after the ``$\cong$'' symbol, consists of a different set of generators for the same stabilizer group, i.e. representing the same stabilizer state. Such alternate generators can be obtained by swapping and multiplying tableau rows (elements from the stabilizer generator), in a process similar to Gaussian elimination~\cite{Audenaert_2005}.

Updating the generators of a stabilizer state after an elementary Clifford gate is applied to the corresponding stabilizer state can be done in time $O(n)$, as follows.
Suppose that $P = \pm P_1 \otimes \dots\otimes P_n$ stabilizes an $n$-qubit state $\ket{\phi}$. Then given an $n$-qubit gate $U$, $UPU^\dagger$ stabilizes $U\ket{\phi}$. This is because $UPU^\dagger U \ket{\phi}= UP\ket{\phi}= U \ket{\phi}$, because $U^{\dagger} U = \unit_{2^n}$ (as $U$ is unitary).
Now if $U$ is a single-qubit operation, we can write $U_j = I \otimes I \otimes \dots \otimes I \otimes U \otimes I \otimes \dots \otimes I$ with $U$ at the $j$-th position in the tensor product.
Therefore, the application of $U$ to the $j$-th qubit of $\ket{\phi}$ updates each element of the stabilizer group to $U_j P U_j^{\dagger} = \pm I P_1 I \otimes \dots \otimes U P_j U^{\dagger} \otimes \dots \otimes IP_n I = \pm P_1 \otimes \dots \otimes U P_j U^{\dagger} \otimes \dots \otimes P_n$.
Since Clifford gates map elements of the Pauli group to elements of the Pauli group, $U P_j U^{\dagger}$ is of the form $\alpha P$ for $P$ a Pauli gate and some $\alpha \in \{\pm 1\}$.\footnote{$\alpha$ cannot be $\pm i$ because $|\alpha|^2 = 1$, which follows from $|\alpha|^2 I = (\alpha P) \cdot (\alpha P)^{\dagger} = (U P_j U^{\dagger}) (U P_j U^{\dagger})^{\dagger} = U P_j U^{\dagger} U P_j^{\dagger} U^{\dagger} = U P_j P_j^{\dagger} U^{\dagger} = U I U^{\dagger} = I$.}
Thus, only the $\pm$ factor in front and $j$-th entry in the tensor product of $P$ should be updated.
This can be done in constant time by a lookup table for each of $H, S$ and each Pauli gate (see \autoref{tab:table}).
Computing the state after applying $H$ or $S$ takes $O(n)$ time in the tableau representation, since we only need to update the $n$ generators of its stabilizer group (a column in the tableau and possibly the column with $\pm$ factors).
A similar procedure works for the two-qubit gate CNOT, also requiring $O(n)$ time to update the stabilizer generators, in this case by modifying two columns of the table and the $\pm$ factors.

\begin{table}[t]
    \caption{Lookup table for the action of conjugating Pauli gates by Clifford gates. The subscripts ``c'' and ``t'' stand for ``control" and ``target".}
    \label{tab:table}
    \centering
    \setlength{\tabcolsep}{12pt} 
    \begin{tabularx}{0.9\textwidth}{c|Xr||c|Xc}
        \toprule
        \textbf{Gate} & \textbf{Input} & \textbf{Output} & \textbf{Gate} & \textbf{Input} & \textbf{Output} \\
        \midrule
        & $X$ & $Z$ & \multirow{6}{*}{CNOT} & $\phantom{-}I_cX_t$ & $\phantom{-}I_cX_t$ \\
        $H$ & $Y$ & $-Y$ & & $\phantom{-}X_cI_t$ & $\phantom{-}X_cX_t$ \\
        & $Z$ & $X$ & & $\phantom{-}I_cY_t$ & $\phantom{-}Z_cY_t$ \\
        \cline{1-3}
         & $X$ & $Y$ & & $\phantom{-}Y_cI_t$ & $\phantom{-}Y_cX_t$ \\
        $S$ & $Y$ & $-X$ & & $\phantom{-}I_cZ_t$ & $\phantom{-}Z_cZ_t$ \\
        & $Z$ & $Z$ & & $\phantom{-}Z_cI_t$ & $\phantom{-}Z_cI_t$ \\
        \bottomrule
    \end{tabularx}
\end{table}

\begin{ex}\label{ex2}
To illustrate the use of the tableau form, we will update the generators $\set{Z_1, Z_2}$ of the $\ket{00}$ state, according to the gates in circuit $A$ of \autoref{ex1}. The ``$\xrightarrow{}$'' symbol will indicate applying one or more gates to the tableau.
\begin{align*}
\set{Z_1, Z_2} = 
\begin{Bmatrix}
Z\otimes I \\
I\otimes Z
\end{Bmatrix}
&\xrightarrow{H_1,H_2}
\begin{Bmatrix}
HZH^{\dagger}\otimes H I H^{\dagger} \\
 H I H^{\dagger}\otimes HZH^{\dagger}
\end{Bmatrix}
= \begin{Bmatrix}
X\otimes I \\
I\otimes X
\end{Bmatrix} \\[2ex] 
&\xrightarrow{CNOT}  
\begin{Bmatrix}
CNOT \ ( X \otimes I ) \ CNOT^{\dagger} \\
CNOT \ ( I \otimes  X ) \  CNOT^{\dagger}
\end{Bmatrix}  
= \begin{Bmatrix}
X\otimes X \\
I\otimes X
\end{Bmatrix} \\[2ex] 
&\xrightarrow{H_1,H_2}  
 \begin{Bmatrix}
H  X  H^{\dagger} \otimes H  X  H^{\dagger} \\
H I H^{\dagger} \otimes H  X  H^{\dagger} 
\end{Bmatrix} 
=  \begin{Bmatrix}
Z\otimes Z \\
I\otimes Z
\end{Bmatrix}
\end{align*}
\end{ex}

\subsection{Circuit Equivalence-Check Problem}

\noindent We proceed to formally state the main problem. We are presented with two $n$-qubit Clifford quantum circuits $U$ and $V$, each represented by (a classical description of) a circuit of only elementary Clifford gates (e.g., $H, S$ and CNOT). The aim of the method is to determine whether or not $U$ and $V$ are equivalent.

\begin{definition}
Fix the number of qubits $n \geq 1$. Given two $n$-qubit unitaries $U, V$, we say that $U$ is \emph{equivalent} to $V$, denoted $U \simeq V$, if $U=cV$ for some complex number $c$.
\end{definition}

The factor $c$ is often called `global phase' and is irrelevant to any observable properties of the two unitaries (for details, see \cite{nielsen2000quantum}).
We remark that if $U=cV$, then $c$ satisfies $|c|^2 = 1$.
This follows from the fact that $U$ and $V$ are unitaries: $\unit = UU^{\dagger} = (cV) \cdot (cV)^{\dagger} = c V \cdot c^* V^{\dagger} = |c|^2 \cdot VV^{\dagger} = |c|^2 \cdot \unit$, hence $|c|^2 = 1$.

\section{Reducing Circuit Equivalence to Classical Simulation}\label{sec:theor_alg}

We explicitly formulate the folkore result (e.g. the two-qubit case is mentioned in \cite[\S 10.5.2]{nielsen2000quantum}) that gives the necessary and sufficient conditions for two unitaries to be equivalent.
We give a self-contained proof which requires minimal prior knowledge.

\begin{theorem}\label{thm:main-theorem}
Let $U, V$ be two unitaries on $n\geq 1$ qubits.
Then $U$ is equivalent to $V$ if and only if the following conditions hold:
\begin{enumerate}
\item for all $j\in \{1, 2, \dots, n\}$, we have $UZ_jU^{\dagger} = V Z_j V^{\dagger}$; and
\item for all $j\in \{1, 2, \dots, n\}$, we have $UX_jU^{\dagger} = V X_j V^{\dagger}$.
\end{enumerate}
Here, as before, we have denoted $Z_j = I \otimes \dots \otimes I \otimes Z \otimes I \otimes \dots  \otimes I$, i.e. an $n$-fold tensor product of identity gates $I$ with the Pauli $Z$ gate at the $j$-th position.
Analogously, $X_j = I \otimes \dots \otimes I \otimes X \otimes I \otimes \dots \otimes I$ where $X$ is the Pauli $X$ gate.
\end{theorem}  
\def\pmpauli{(\pm 1)\textnormal{-Pauli}\xspace}
\begin{proof}
If $U\simeq V$, then $U=cV$ for some $c\in \mathbb{C}, |c|=1$, so $U Z_j U^{\dagger} = c V Z_j (cV)^{\dagger} =c V Z_j (V)^{\dagger} \cdot c^* = |c|^2 V Z_j V^{\dagger} = V Z_j V^{\dagger}$ and similarly for $X_j$ where $c^*$ is the complex conjugate of $c$.

For the converse direction, we first note that if $U$ and $V$ coincide on $X_j$ and $Z_j$ by conjugation, then they must coincide by conjugation on \emph{all Pauli strings}.
The reason for this is that any Pauli string can be written as a product of $\{X_j, Z_j\}_{j=1}^n$ modulo a complex number from $\{\pm 1, \pm i\}$.
Given such a product $P = \prod_{k=1}^n X_k^{x_k} Z_k^{z_k}$ where $x_k, z_k \in \{0, 1\}$ determine if $X_k$ or $Z_k$ is included in the product, we see that $
UPU^{\dagger}
=
U\left(\prod_{k=1}^n X_k^{x_k} Z_k^{z_k}\right) U^{\dagger} 
=
\prod_{k=1}^n UX_k^{x_k}U^{\dagger} U Z_k^{z_k} U^{\dagger} 
$.
This shows that $UPU^{\dagger} = VPV^{\dagger}$ if $UX_kU^{\dagger} = VX_k V^{\dagger}$ and $UZ_kU^{\dagger} = VZ_k V^{\dagger}$ for all $k=1,2,\dots, n$.

Given an $n$-qubit quantum state $\ket{\phi}$, we can write
	\begin{equation}
\dyad{\phi} = \sum_{
P
} \alpha_P P
		\label{eq:phi-in-pauli-basis}
	\end{equation}
	where the summation runs over all Pauli strings of length $n$, i.e. $P \in \textsc{Pauli}_n = \{I, X, Y, Z\}^{\otimes n}$, and the weights $\alpha_P \in \mathbb{C}$ are unique, i.e. each $\textsc{Pauli}_n$ string $P$ is associated with a weight $\alpha_P$.
	The reason this can be done is that any $2^n \times 2^n$ matrix with complex entries can be written as linear combination of $n$-qubit Pauli strings \cite{nielsen2000quantum,KIMURA2003339} (the `Pauli basis') and $\dyad{\phi}$ indeed is an $2^n \times 2^n$ matrix.
	Now by conjugating both sides of \autoref{eq:phi-in-pauli-basis}, we obtain
\begin{equation}
U \dyad{\phi} U^{\dagger} = \sum_{P} \alpha_P UPU^{\dagger}
\end{equation}
	which by the observation above (that $U$ and $V$ coincide on all Pauli strings by conjugation) equals
\begin{equation}
V \dyad{\phi} V^{\dagger} = \sum_{P} \alpha_P VPV^{\dagger}
\end{equation}
hence
	\begin{equation}
		\label{eq:a-on-phi}
	A \dyad{\phi} A^{\dagger} = \dyad{\phi}
\end{equation}
where $A=V^{\dagger} U$.
	This arises from the fact that $A^{\dagger}= (V^{\dagger}U)^{\dagger}=U^{\dagger}V$ and, since $U$ and $V$ coincide on all Pauli strings by conjugation, \begin{align}
A \dyad{\phi} A^{\dagger} &= V^{\dagger} U \dyad{\phi} U^{\dagger} V = V^{\dagger} \Big( \sum_{P} \alpha_P UPU^{\dagger} \Big) V^{\dagger} \notag \\
&= V^{\dagger} \Big( \sum_{P} \alpha_P VPV^{\dagger} \Big) V = \dyad{\phi}
\end{align}
since $V^{\dagger}$ cancels out with $V$. 
 
	By applying $\bra{\phi}$ from the left and $\ket{\phi}$ to the right on both sides of \autoref{eq:a-on-phi}, we obtain $|\expval{A}{\phi}|^2 = |\langle \phi | \phi \rangle|^2 = 1$.
Thus, the modulus of the inner product between $A\ket{\phi}$ and $\ket{\phi}$ equals the product of their norms (which both equal 1), hence the tightness condition of the Cauchy-Schwarz inequality implies that $A\ket{\phi}$ and $\ket{\phi}$ are linearly dependent.
That is, $\ket{\phi}$ is an eigenvector of $A$.

Since this holds for arbitrary $n$-qubit states $\ket{\phi}$, each vector is an eigenvector of $A$.
By standard linear algebra, we know that this implies that $A$ is a multiple of the identity operator.
	Thus $A = c \unit_{2^n}$ for some complex number $c$, hence $V^{\dagger} U = c\unit_{2^n}$ by definition of $A$.
	Applying $V$ to the left of both sides of $V^{\dagger} U = c\unit_{2^n}$ yields $U = cV$.
\qed
\end{proof}

The above theorem is applicable to general quantum circuits. However, for the purposes of this study, we will concentrate on Clifford circuits. In that case, the theorem induces an algorithm which reduces the equivalence checks in Part 1 and 2 of the theorem to simulating the circuits $U$ and $V$, which for the case of Clifford circuits, is well known to be efficient~\cite{gottesman1998heisenberg,aaronson2008improved}.

\paragraph{\textbf{The algorithm.} } \label{algorithm}
\noindent From \autoref{sec:preliminaries}, we know that $S_0 = \{Z_j \mid j=1,2,\dots, n \}$ generate the stabilizer group of the state $\ket{0}^{\otimes n}$, and thus $S_0$ ``represents'' $\ket{0}^{\otimes n}$ in the stabilizer formalism.
The same holds for $\{X_j \mid j=1,2,\dots, n\}$ and the state $\ket{+}^{\otimes n}$, where $\ket{+} = \frac{1}{\sqrt{2}} \left(\ket{0} + \ket{1}\right)$.
Furthermore, updating a stabilizer state representation $\{g_1, g_2, \dots, g_n\}$ (i.e. the $g_j$ are generators of the state's stabilizer group), after a Clifford gate $U$ is found as $\{Ug_1 U^{\dagger}, \dots, Ug_n U^{\dagger}\}$.
So computing $U Z_j U^{\dagger}$ for all $j = 1, 2, \dots, n$ is the same as computing the classical simulation of $U \ket{0}^{\otimes n}$ in the stabilizer formalism and computing $U X_j U^{\dagger}$ for all $j = 1, 2, \dots, n$ is the same as classical simulation of $U \ket{+}^{\otimes n}$.
Combining these facts with \autoref{thm:main-theorem}, we obtain the following algorithm for equivalence checking of Clifford circuits which is essentially a (many-one) reduction to Clifford circuit simulation.

To be explicit, given Clifford circuits $U, V$, we determine whether they are equivalent as follows:
\begin{enumerate}
\item Simulate $U$ gate-by-gate in the stabilizer formalism, where the stabilizer group generators of the input state are $\{Z_1, Z_2, \dots, Z_n\}$, i.e. the input state is $\ket{0}^{\otimes n}$. This yields the output generator set $\{UZ_1U^{\dagger}, UZ_2U^{\dagger}, \dots, UZ_nU^{\dagger}\}$.
\item Do the same for $V$, yielding $\{VZ_1V^{\dagger}, VZ_2V^{\dagger}, \dots, VZ_nV^{\dagger}\}$.
\item Check for each $j=1, 2, \dots, n$, whether the Pauli elements $UZ_j U^{\dagger}$ and $VZ_jV^{\dagger}$ are equal. If there is some $j$ for which they are non-equal, return ``Non-equivalent.''
\item Repeat steps (1-3) for the input stabilizer generator set $\{X_1, X_2, \dots, X_n\}$, which is produced by starting with the generator set of $\ket{0}^{\otimes n}$, followed by applying the Hadamard gate $H$ on each qubit (since $HZH^{\dagger} = X$).
\item If the algorithm reaches this point, $U$ and $V$ agree by conjugation on all $X_j$ and $Z_j$. Return ``Equivalent.''
\end{enumerate}

\autoref{ex3} shows an example execution of the algorithm on the two-qubit circuits we saw before.

\begin{ex}\label{ex3}

	\noindent We aim to determine whether circuits $A$ and $B$ from \autoref{ex1}
	are equivalent.
	Following the algorithm above, we need to compute the output, under conjugation, of the two circuits for each of the inputs $Z_1, Z_2, X_1, X_2$.

	\begin{enumerate}[a)]

\item Circuit $A$ on $\set{Z_1,Z_2}$:\\
 \autoref{ex2} shows that the resulting tableau is $\begin{Bmatrix}
Z\otimes Z \\
I\otimes Z
\end{Bmatrix}$\\

\item Circuit $B$ on $\set{Z_1,Z_2}$:
\begin{align*}
\begin{Bmatrix}
Z\otimes I \\
I\otimes Z
\end{Bmatrix}
&\xrightarrow{NOTC}
\begin{Bmatrix}
NOTC \ ( Z \otimes I ) \ NOTC^{\dagger} \\
NOTC \ (I \otimes Z ) \ NOTC^{\dagger}
\end{Bmatrix}  = \begin{Bmatrix}
Z\otimes Z \\
I\otimes Z
\end{Bmatrix} 
\end{align*}

\item Circuit $A$ on $\set{X_1,X_2}$:
\begin{align*}
\begin{Bmatrix}
X\otimes I \\
I\otimes X
\end{Bmatrix}
&\xrightarrow{H_1,H_2}
\begin{Bmatrix}
HXH^{\dagger}\otimes H I H^{\dagger} \\
 H I H^{\dagger}\otimes HXH^{\dagger}
\end{Bmatrix}
= \begin{Bmatrix}
Z\otimes I \\
I\otimes Z
\end{Bmatrix} \\[2ex]
&\xrightarrow{CNOT}   \begin{Bmatrix}
CNOT \ ( Z \otimes I ) \ CNOT^{\dagger} \\
CNOT \ ( I \otimes Z ) \ CNOT^{\dagger}
\end{Bmatrix}  = \begin{Bmatrix}
Z\otimes I \\
Z\otimes Z
\end{Bmatrix} \\[2ex] 
&\xrightarrow{H_1,H_2}  \begin{Bmatrix}
H  Z  H^{\dagger} \otimes H  I  H^{\dagger} \\
H Z H^{\dagger} \otimes H  Z  H^{\dagger} 
\end{Bmatrix}  =  \begin{Bmatrix}
X\otimes I \\
X\otimes X
\end{Bmatrix} 
\end{align*}

\item Circuit $B$ on $\set{X_1,X_2}$:
\begin{align*}
\begin{Bmatrix}
X\otimes I \\
I\otimes X
\end{Bmatrix}
&\xrightarrow{NOTC}
\begin{Bmatrix}
NOTC \ ( X \otimes  I ) \ NOTC^{\dagger} \\
NOTC \ ( I \otimes  X ) \ NOTC^{\dagger}
\end{Bmatrix}  = \begin{Bmatrix}
X\otimes I \\
X\otimes X
\end{Bmatrix} 
\end{align*}
\end{enumerate}

We observe that the two circuits output, under conjugation, the same elements of the Pauli group, for each of $Z_1$, $Z_2$, $X_1$ and $X_2$.
Therefore, we can conclude that circuits $A$ and $B$ are equivalent.

\end{ex}

We note that requiring that $U$ and $V$ agree on all $X_j$ and $Z_j$ by conjugation (step 3 of the Algorithm) is a stronger statement than requiring that $U$ and $V$ output the same state on input $\ket{0}^{\otimes n}$ and $\ket{+}^{\otimes n}$.
As counterexample for $n=2$, 
consider the identity circuit $\unit_4$. We have 
$\unit_4 \ket{00} = \ket{00} = U_A \ket{00} = U_B \ket{00}$ and
$\unit_4 \ket{++} = \ket{++} = U_A \ket{++} = U_B \ket{++}$, where $U_A$ and $U_B$ are the unitaries for circuits $A$ and $B$ from \autoref{ex1}, respectively.
This calculation can also be done in terms of stabilizer tableaux using the $\cong$ relation to check if the tableaux generate the same stabilizer group:
$$
~~\underbrace{\begin{Bmatrix}
Z\otimes I \\
I\otimes Z
\end{Bmatrix}}_{\text{for }\unit_4 \ket{00}}
~~\cong\hspace{-.5em}
\underbrace{\begin{Bmatrix}
I\otimes Z \\
Z\otimes Z
\end{Bmatrix}}_{\text{for }U_A \ket{00} = U_B \ket{00}}
  \text{  and  ~~~ } 
\underbrace{\begin{Bmatrix}
X\otimes I \\
I\otimes X
\end{Bmatrix} }_{\text{for }\unit_4 \ket{++}}
~~\cong\hspace{-.5em}
\underbrace{\begin{Bmatrix}
X\otimes I \\
X\otimes X
\end{Bmatrix}}_{\text{for }U_A \ket{++} = U_B \ket{++}}
$$
That is, the list of generators of the stabilizer groups of $\unit_4\ket{00}$ ($\unit_4\ket{++}$) and $U_A\ket{00} = U_B \ket{00}$ ($U_A\ket{++} = U_B\ket{++}$) are \emph{equivalent} in the sense that they give rise to the same stabilizer group (and hence represent the same quantum state) but they are not \emph{equal}.
Unitaries are only equivalent if they output equal Pauli group elements under conjugation.
Indeed, $\unit_4$ is not equivalent to $U_A$ or $U_B$: a witness to their non-equivalence is the state $\ket{01}$, since $\unit_4 \ket{01} =\ket{01} \neq \ket{11} = U_A \ket{01}  =  U_B \ket{01}$.

Since storing the $n$ stabilizer generators for an $n$-qubit state requires $O(n^2)$ space, naively initializing the tableaux $\{Z_1, \dots, Z_n\}$ and $\{X_1, \dots, X_n\}$ takes $O(n^2)$ time.
Next, updating a tableau for a single-qubit gate or two-qubit gate takes time $O(n)$.
Hence the runtime of the algorithm is $O(n^2 + m\cdot n)$ with $m$ the sum of the number of elementary Clifford gates in $U$ and $V$.
However, we can avoid the $O(n^2)$ initialization time, by amortizing the creation of the stabilizer generator set over the update operations using a lazy initialization approach.
To be precise, let us keep track of the stabilizer generators by listing them in an $n \times (n+1)$ matrix where the $(n+1)$-th column consists of factors $\pm 1$ and the entries of the first $n$ columns are Pauli gates~\cite{Garcia2014}.
Instead of initializing this matrix, we only mark all columns as uninitialized.
The uninitialized entries in column $k$ of the matrix are filled when a gate is applied to the $k$-th qubit for the first time, in which case the algorithm runs over all elements of column $k$ anyway (and the column of $\pm$ factors).
This lazy initialization brings the total runtime of the algorithm down to $O(m\cdot n)$.

\section{Experiments}\label{sec:experiments}

We implemented the algorithm from \autoref{sec:theor_alg} in Python using the open-source Stim Clifford-circuit simulator~\cite{Gidney2021stimfaststabilizer} as a backend.
See \cite{githubrepo} for our open-source implementation.

We empirically evaluated the implementation and compared the runtime to QuSAT~\cite{berent2022towards,qusatgithubrepo}, a recent SAT-based Clifford equivalence checker and
the Feynman tool~\cite{amy2018towards}.
We used a laptop with a 3.2 GHz M1 processor with 8Gb RAM.

To make a fair comparison, we adopt the experimental setting of \cite{berent2022towards} and generate random circuits using QuSAT, which consists of generating random sequences of elementary Clifford gates $H, S, \textnormal{CNOT}$.
QuSAT generates circuits which are completely `filled' in the sense that if the depth is $d$, the number of gates applied to each qubit is also $d$; thus, since only $H, S, \textnormal{CNOT}$ are used, the number of gates in a depth-$d$ circuit is between $d\cdot \frac{n}{2}$ (only two-qubit gates) and $d\cdot n$ (only single-qubit gates).
We emphasize that the runtime of this work's method is deterministic and a function of the number of qubits and the number of gates only, and is hence independent of the gates that the two input circuits consist of, or the ordering of the gates.

We validated the correctness of our implementation by comparing its results with QuSAT. Across all circuit pairs in which QuSAT terminated, we found a consistent classification as either equivalent or non-equivalent by both approaches.

\begin{figure}
  \begin{subfigure}[b]{0.495\textwidth}
    \captionsetup{width=.8\linewidth}
    \includegraphics[width=\textwidth]{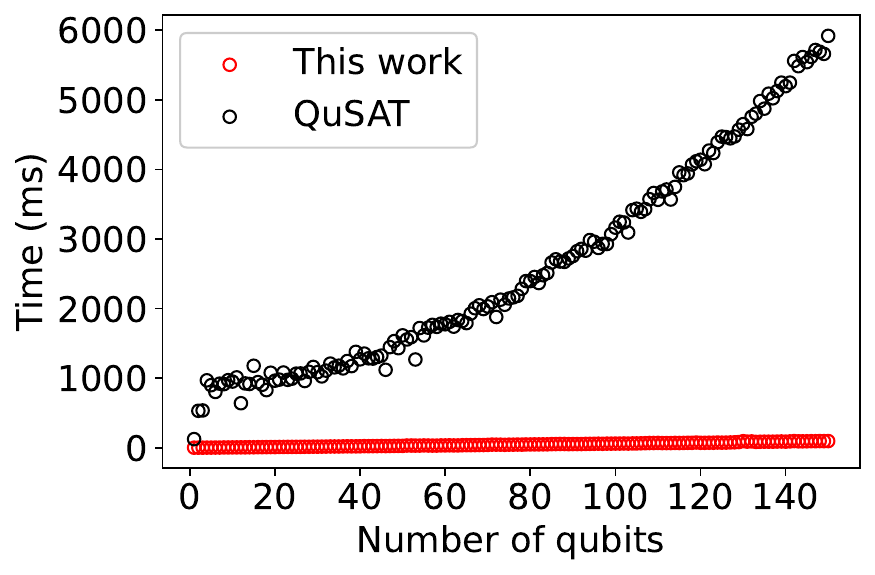}
    \caption{
      \small Fixed depth of 1000. This data-set reappears in \autoref{fig:plotsc}.
      \vspace{6.3\baselineskip}
    }\label{fig:plotsa}
  \end{subfigure}
  \centering
  \begin{subfigure}[b]{0.49\textwidth}
    \captionsetup{width=.9\linewidth}
    \includegraphics[width=\textwidth]{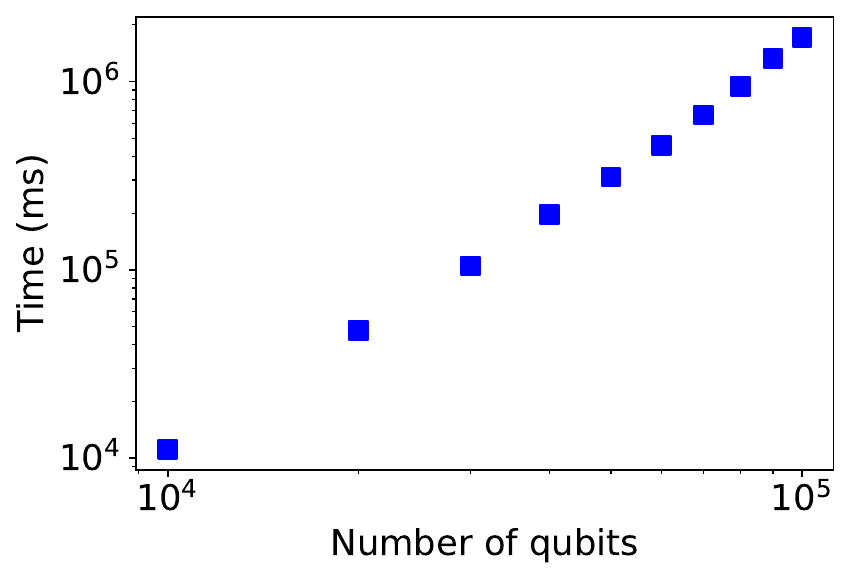}
    \caption{
      \small The approach from this work, reaching beyond what was previously feasible: Fixed depth 10 and the number of qubits up to 100.000. Both axes are in logarithmic scale, so that a polynomial scaling shows up as a straight line.
      \vspace{1.2\baselineskip}
    }\label{fig:plotsb}
  \end{subfigure}
  \begin{subfigure}[b]{0.495\textwidth}
    \captionsetup{width=.9\linewidth}
    \includegraphics[width=\textwidth]{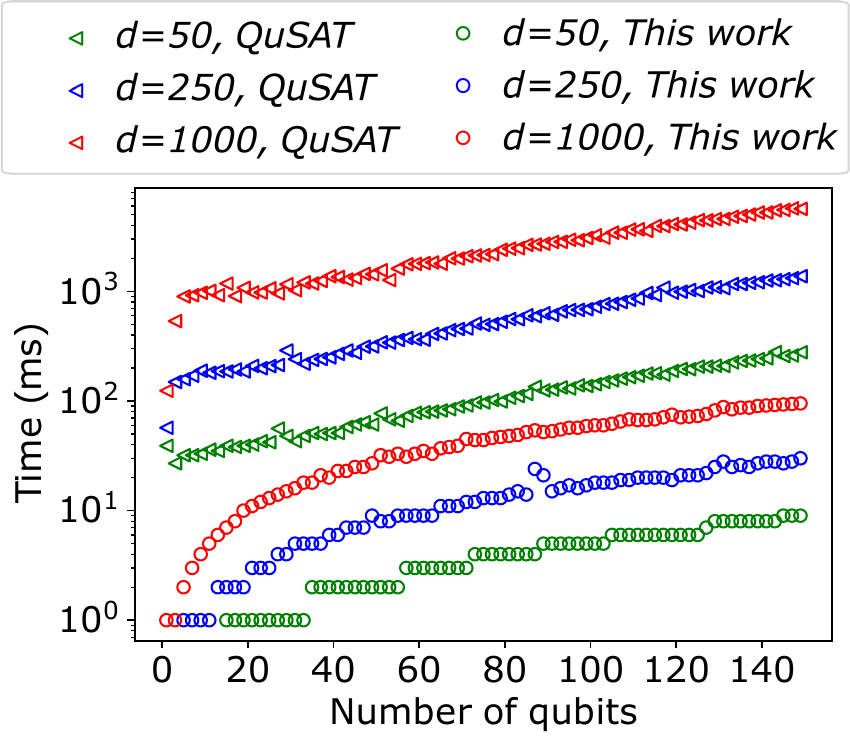}
    \caption{
      Fixed depths (``d") and increasing number of qubits. The vertical axis is on logarithmic scale.
      \vspace{2\baselineskip}
    }\label{fig:plotsc}
  \end{subfigure}
  \hfill
  \begin{subfigure}[b]{0.495\textwidth}
    \captionsetup{width=.9\linewidth}
    \includegraphics[width=\textwidth]{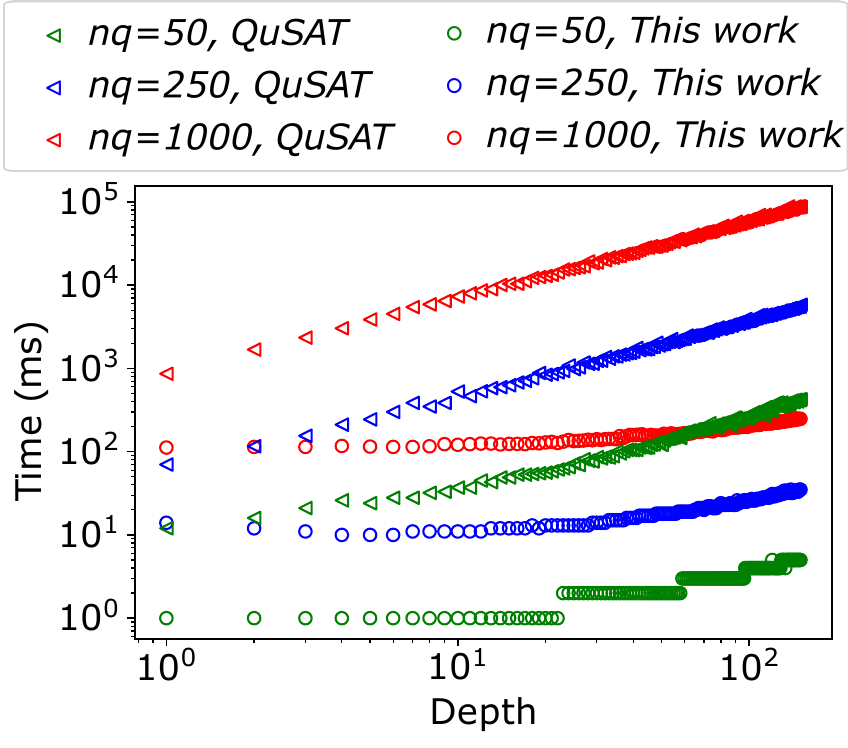}
    \caption{
      \small Fixed number of qubits (``nq") and increasing depth. Both axes are on logarithmic scale, so that a polynomial scaling shows up as a straight line.
      \vspace{1\baselineskip}
    }\label{fig:plotsd}
  \end{subfigure}
  \caption{
    \small Runtime comparison of circuit-equivalence checking between equivalent randomly-generated Clifford circuits, for both the method from this work and QuSAT. The step pattern observed for lower values is a result of the limitations of the time-measuring function which operates in milliseconds. The runtimes for non-equivalent circuits (not displayed) have an indistinguishable appearance.
  }\label{fig:plotsabc}
\end{figure}

First, we ran both QuSAT and our implementation on both equivalent and non-equivalent random Clifford circuits which were thus produced by QuSAT.
The results are shown in \autoref{fig:plotsabc}, for varying number of qubits (\autoref{fig:plotsa}, \autoref{fig:plotsb} and \autoref{fig:plotsc}) and varying quantum-circuit depth (\autoref{fig:plotsd}).
We performed an equal amount of experiments with non-equivalent circuit pairs, again drawing them from the experimental setting of QuSAT~\cite{berent2022towards}. The resulting plots appear indistinguishable from the ones for equivalent circuit pairs, and for that reason we omit them here.

We observe that the implementation is very fast and can handle large circuits: up to $1000$ qubits with a depth of $10.000$ gates in $\sim$22 seconds, and $100.000$ qubits with $10$ gates in $\sim$15 minutes (\autoref{fig:plotsb}).
The tested regime of the method consistently outperforms QuSAT by one to two orders of magnitude ($10\times$ to $100\times$) or even more.
We also see that the runtime of QuSAT, whose runtime is heuristic, seems to scale exponentially in the number of qubits whereas the runtime of our approach is deterministic and scales polynomially in both number of qubits and number of gates (\autoref{sec:theor_alg}).

\begin{figure} 
  \begin{subfigure}[b]{0.495\textwidth}
    \captionsetup{width=.8\linewidth}
    \includegraphics[width=\textwidth]{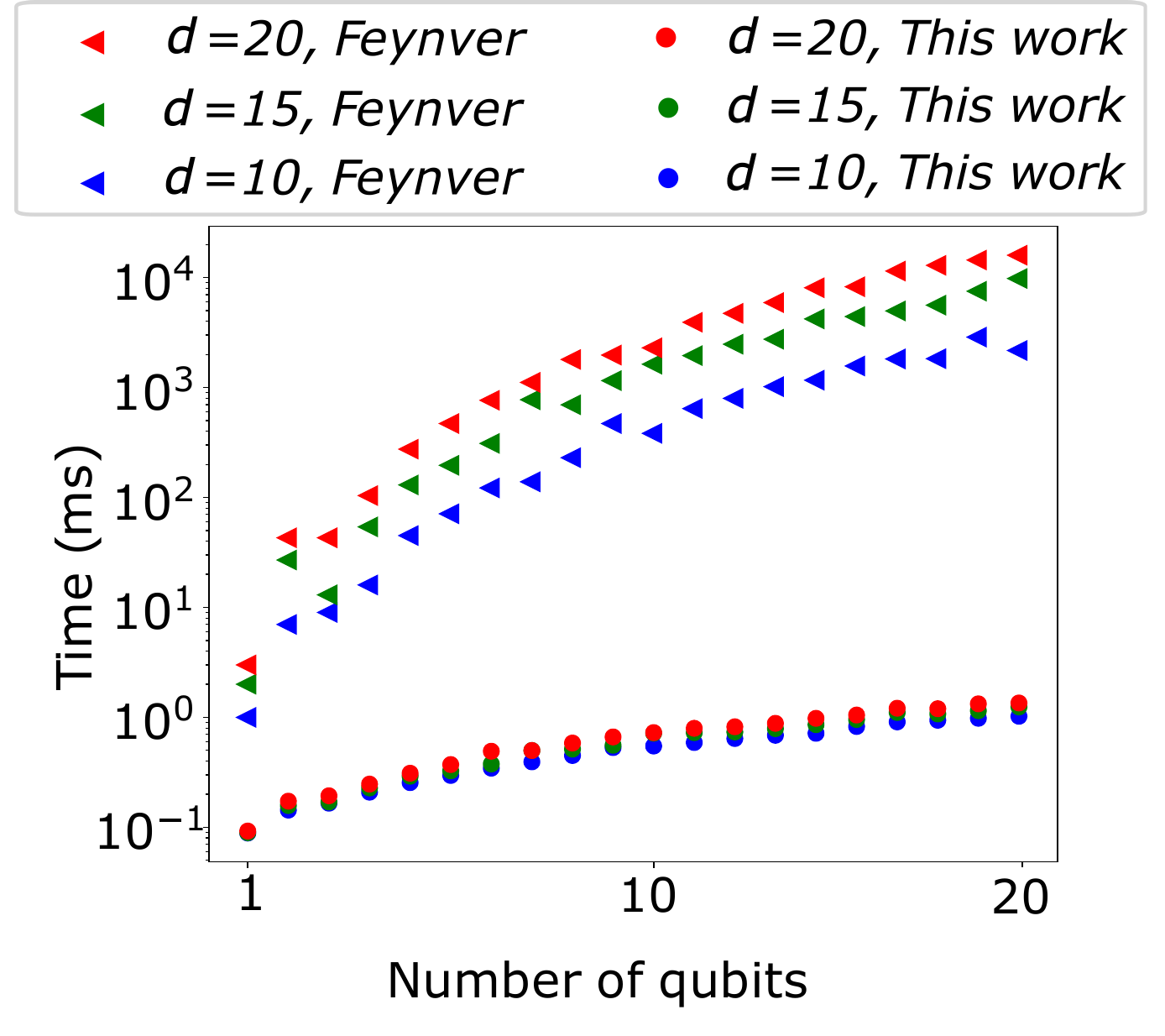}
    \caption{
      \small Fixed depths (``d") and increasing number of qubits. The vertical axis is on logarithmic scale.
      \vspace{1.3\baselineskip}
    }\label{feynv:a}
  \end{subfigure}
  \centering
  \begin{subfigure}[b]{0.495\textwidth}
    \captionsetup{width=.9\linewidth}
    \includegraphics[width=\textwidth]{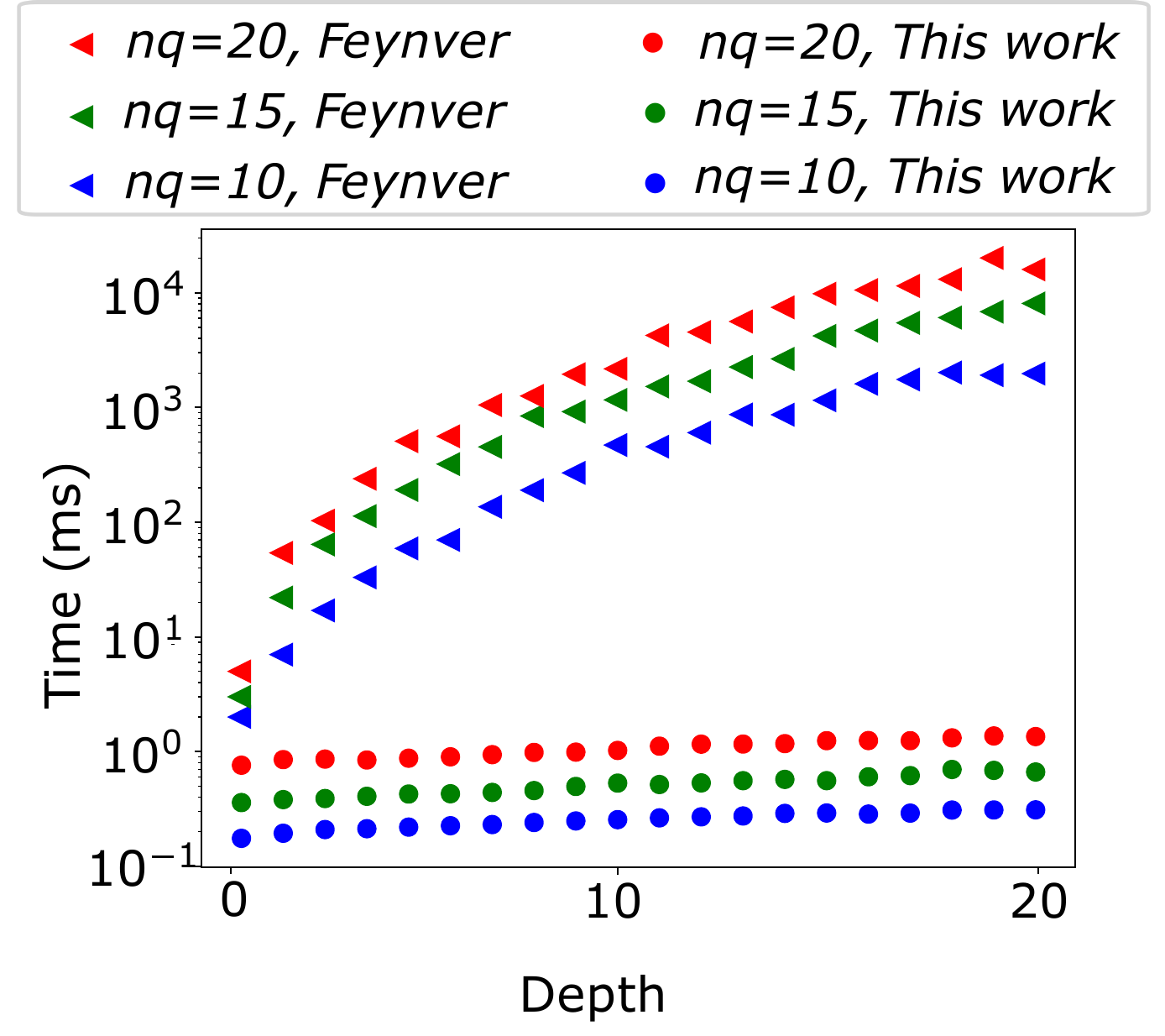}
    \caption{
      \small Fixed number of qubits (``nq") and increasing depth. The vertical axis is on logarithmic scale.
      \vspace{2.4\baselineskip}
    }\label{feynv:b}
  \end{subfigure}
  \caption{
    \small Runtime of circuit-equivalence checking between equivalent randomly-generated Clifford circuits for various circuit depths and the number of qubits.
  }\label{fig:feyn}
\end{figure}

Next, we also compare with the Feynver submodule for circuit-equivalence checking of the Feynman tool~\cite{amy2018towards}.
Feynver is based on Feynman path-integrals and it can verify the equivalence of general quantum circuits.
For Clifford circuits specifically, its runtime scales polynomially in the number of elementary gates for Clifford circuits, just like the method presented in this work.
To compare to Feynver, we again let QuSAT generate random circuits and input them to both Feynver and our implementation.
We found that Feynver terminated on all equivalent circuit pairs, but for non-equivalent pairs it often aborted. This behavior is known~\cite{amy}.
Our experiments showed that Feynver is outperformed by both QuSAT and our implementation.
For equivalent circuit pairs, the runtimes of Feynver and our implementation are shown in \autoref{fig:feyn} for a varying number of qubits (\autoref{feynv:a}) and a varying quantum-circuit depth (\autoref{feynv:b}). The running times for the non-equivalent pairs for which Feynver terminated successfully were similar.

\section{Conclusions}\label{sec:conclusions}
 
In this paper, we demonstrate that a deterministic algorithm, which is based on a folklore mathematical result, can surpass the efficiency of current methods for exact equivalence checking of quantum circuits consisting of Clifford gates. The algorithm reduces equivalence checking to classical simulation of Clifford circuits and runs in time $O(n \cdot m)$, with $n$ the number of qubits and $m$ the total number of elementary Clifford gates of the two input circuits.
This scaling implies efficient equivalence checking for various application-relevant circuits, for example the circuits for producing the two-dimensional cluster states (resource states for universal quantum computing~\cite{briegel2009measurement}), GHZ states (resource states for various quantum communication protocols~\cite{hein2006entanglement}) or performing error detection (excluding the measurement) in quantum error correction~\cite{terhal2015quantum}.

We have implemented the algorithm using the Stim simulator and tested it on a variety of benchmark circuits with different sizes and depths, and compared it to two existing state-of-the-art methods, QuSAT and Feynver.
Our results, reaching $1000$ qubits (with depth $10.000$) in less than a minute and $100.000$ qubits (depth $10$) in $\sim$15 minutes, demonstrate that this approach consistently outperforms both.
Furthermore, since the method is deterministic, its scaling behavior is known.

Possible future work includes extending this method to arbitrary circuits using non-Clifford gates~\cite{arunachalam2022parameterized,amy2018towards}, following existing classical simulation formalisms of such circuits~\cite{bravyi2018simulation}. Our preliminary results towards this direction appear promising.  

\section{Acknowledgments}
We thank Prof. M. Bonsangue for helpful discussions.
This work was supported by the Dutch National Growth Fund, as part of the Quantum Delta NL program.

\bibliographystyle{llncs2e/splncs03}
\bibliography{lit}

\end{document}